# Antichiral hinge states in a higher-order photonic nodal ring semimetal

Yuchen Peng[1,#], Deep Mondal[2,#], Yuting Yang[3,*], Wanting Wu[3], Bo Zhao[1], Shuaiyang Wei[1], Zhenzhi Liu[1], Wenrong Qi[4], Xiaokang Dai[1], Minqi Cheng[1], Weili Li[1], Xinyi Zhang[1], Fei-Fei Li[5], Rimi Banerjee[6], Minggui Wei[7], Jingyi Tian[1], Peiheng Zhou[8], Subhaskar Mandal[2,*], Baile Zhang[7,*], Gui-Geng Liu[1,*]

[1]*Department of Electronic and Information Engineering, School of Engineering, Westlake University, Hangzhou 310030, China*
[2]*Department of Physics, Indian Institute of Technology Bombay, Mumbai 400076, India*
[3]*School of Materials Science and Physics, China University of Mining and Technology, Xuzhou 221116, China*
[4]*School of Physics, Henan Normal University, Xinxiang 453007, China*
[5]*School of Information and Control Engineering, China University of Mining and Technology, Xuzhou 221116, China*
[6]*Department of Physics, Indian Institute of Technology Dharwad, Dharwad 580011, India*
[7]*Division of Physics and Applied Physics, School of Physical and Mathematical Sciences, Nanyang Technological University, Singapore 637371, Singapore*
[8]*National Engineering Research Center of Electromagnetic Radiation Control Materials, State Key Laboratory of Electronic Thin Film and Integrated Devices, University of Electronic Science and Technology of China, Chengdu 610054, China*

[#]*These authors contributed equally to this work.*
[*]*Correspondence: yangyt@cumt.edu.cn (Y. Yang); subhaskar@iitb.ac.in (S. Mandal); blzhang@ntu.edu.sg (B. Zhang); liuguigeng@westlake.edu.cn (G.-G. Liu)*

**Antichiral states propagate in the same direction on opposite boundaries, defying the conventional constraint that boundary modes must cancel net chirality. Previously found antichiral states have been limited to first-order topological semimetals. However, antichiral hinge states—the antichiral counterpart of recently discovered higher-order chiral hinge states—remain elusive. Here, we report the observation of antichiral hinge states in a higher-order nodal-ring semimetal made of a three-dimensional gyromagnetic photonic crystal. Near-field scanning measurements reveal a pair of hinge states at two parallel one-dimensional boundaries propagating unidirectionally along the same direction, with robust transport against metallic scatterers. Their spatial positions can be reconfigured by adding or removing photonic layers. The additionally observed antichiral and drumhead surface states manifest a hierarchy of first- and second-order topological boundary states within a single photonic system. Our work extends antichiral states to higher-order topological semimetals and has potential applications in robust and reconfigurable photonic routing.**

Chiral transport, characterized by unidirectional propagation robust against backscattering, is central to topological band theory and is the most direct manifestation of nontrivial bulk topology [1,2]. Chiral transport was first observed as one-dimensional (1D) chiral edge states in two-dimensional (2D) systems, initially in the quantum Hall effect [3] and subsequently in Chern insulators [4] and their classical analogues [5,6]. Extending this principle to three dimensions (3D), 2D chiral surface states have been observed in 3D quantum Hall systems [7,8] and 3D Chern insulators [9,10]. In all those gapped insulators, boundary modes on opposite parallel sides propagate in opposite directions [1,2]. This arises from a fundamental constraint: on a closed manifold, the net chirality, defined as the algebraic sum of left- and right-moving boundary channels, must vanish, thereby forcing oppositely directed modes on opposing boundaries [1,2,11].

On the other hand, antichiral boundary states can circumvent this constraint by displacing the required counter-propagating modes into the gapless bulk [12-16]. In these systems, the bulk is a gapless semimetal, and boundary modes on opposite sides propagate in the same direction. This concept was first proposed in a modified Haldane model [12] and subsequently realized in 2D Dirac semimetals as antichiral edge states [13,14] and in 3D Weyl semimetals as antichiral surface states [15,16]. All the above findings, whether chiral or antichiral, belong to first-order topology, where the boundary modes are exactly one dimension lower than the bulk.

The recent observation of 1D chiral hinge states in a 3D photonic axion insulator [17,18] extended unidirectional transport into higher-order topology [19-21]. This raises a natural question: can antichiral hinge states, the higher-order counterpart, be realized? Note that in 3D higher-order real materials (without synthetic dimensions), the 1D second-order hinge state is the only possible propagating boundary channel (third-order corner states are localized), making antichiral hinge states the last possible missing class of unidirectional transport modes [20]. Unlike chiral hinge states emerging in gapped axion insulators, antichiral hinge states require a gapless bulk and thus can only be realized in higher-order semimetals. Recently, theoretical proposals for antichiral hinge states have also been put forward based on coupled Haldane models [22] and photonic synthetic dimensions [23]. Despite tremendous efforts made in the finding of higher-order topological materials [19-21], no observation of antichiral hinge states has been reported.

In this work, we theoretically propose and experimentally demonstrate antichiral hinge states in a 3D higher-order photonic nodal ring semimetal. Our platform is a 3D gyromagnetic photonic crystal that hosts a pair of nodal rings and two pairs of Dirac lines with shifted frequencies. Through near-field scanning measurements, we observe flat drumhead surface states arising from the projection of the nodal rings, as well as dispersive antichiral surface states connecting projected Dirac lines. A pair of antichiral hinge states is observed at two parallel 1D boundaries. Arising from the surface polarization of the antichiral surface states, the two hinge states propagate unidirectionally in the same direction and remain robust even in the presence of metallic obstacles. Moreover, the spatial positions of these hinge states can be reconfigured simply by adding or removing photonic crystal layers, enabling on-demand control over which hinges host the topological modes. Our results extend antichiral physics to higher-order topology and offer a new route toward tunable chiral transport.

## Results

### Topological hierarchy and multiple degeneracy lines in a 3D magnetic photonic crystal

Figure 1a depicts the unit cell of our designed photonic crystal, which consists of stacked two-dimensional gyromagnetic photonic layers isolated by metallic coupling plates. Within each photonic layer, two Yttrium Iron Garnet (YIG) rods (gyromagnetic materials) are positioned at the corners, biased by opposite magnetic fields via permanent magnets, whose configuration is similar to that used for realizing antichiral edge states [13,24]. Two types of alternating metallic plates are employed to introduce interlayer coupling between adjacent photonic layers. Each plate features coupling holes that break in-plane inversion symmetry while preserving $C_{3v}$ symmetry, and the hole patterns on the two plate types are inversion-symmetric with respect to each other. This photonic structure can be effectively described by the tight-binding model shown in Fig. 1b, which consists of stacked typical modified Haldane layers [12] with alternating vertical interlayer couplings (similar to previous work in 3D Chern insulator [9] and axion insulator [17]). Note that the interlayer couplings to the two sublattices differ within a single layer (see Supplementary Note 1). Compared with previous theoretical proposals [22,23], our 3D real-space model provides the first route to directly observe intrinsic antichiral hinge states, where the required counter-propagating compensation is carried by gapless bulk modes rather than by boundary modes such as internal-interface states [22].

Figure 1c plots the calculated bulk band dispersion of the photonic crystal, revealing a semimetal phase formed by four bands. Along each of the high-symmetry lines K–H and K′–H′, two Dirac lines appear: one between the first and second bands and the other between the third and fourth bands, as indicated by the green lines in Fig. 1c, d. The Dirac lines along K–H occur at higher frequencies than their counterparts along K′–H′. As illustrated in Fig. 1e, f, the bands near the Dirac line exhibit linear dispersion in the $k_x$-$k_z$ plane while remaining degenerate along the $k_z$ direction (green lines); note that the bands become degenerate throughout the Brillouin zone (BZ) at $k_z$=±π/$h$. In addition, two nodal rings are formed between the second and third bands, encircling the K and K′ points in the $k_x$-$k_y$ plane (pink and yellow circles in Fig. 1g, h, respectively). Each nodal ring carries a Berry phase of π when integrated along a closed loop encircling it [21]. The vertical Dirac lines are also characterized by a quantized π Berry phase (see detailed derivations in Supplementary Note 1). The dispersion along the nodal ring is linear in the radial direction (perpendicular to the ring) and nearly flat along the tangential direction (following the ring), as shown in Fig. 1g, h. The two nodal rings around K and K′ exhibit a substantial frequency shift: the ring around K lies at approximately 8.8 GHz, while that around K′ is at around 7.0 GHz. As illustrated in Fig. 1d, the momentum-space geometry shows that each vertical Dirac line threads the interior of a closed nodal ring without crossing the ring itself, forming a linked line–ring configuration of bulk degeneracies. Furthermore, we directly mapped these bulk Dirac lines and nodal rings using momentum-resolved bulk-field measurements. The reconstructed three-dimensional Fourier spectra resolve line-like degeneracies extending along the vertical momentum direction near the K and K′ valleys and ring-shaped contours at the corresponding nodal-ring frequencies, confirming the three-dimensional semimetal phase (see Supplementary Note 2).

As illustrated in Fig. 1i–l, the bulk properties of our photonic crystal give rise to a hierarchy of multidimensional topologically protected boundary states, including the drumhead surface states on the bottom or top surfaces ($z$-normal) (Fig. 1j), antichiral surface states on the front and back surfaces ($y$-normal) (Fig. 1l), and higher-order antichiral hinge states at the intersection between the $z$-normal and $y$-normal boundaries (Fig. 1k).

**Experimental observation of two distinct types of topological surface states**

Our fabricated sample is shown in Fig. 2a–d. The metallic copper plates feature two distinct boundary terminations along the $y$-direction, labelled type 1 and type 2 in Fig. 2c, d. To probe the drumhead surface states, we cover the top surface with copper cladding while wrapping the others with microwave absorbers. A source antenna connecting to the microwave vector network analyzer is placed near the top boundary, and the excited fields are mapped using a near-field scanning setup (see Supplementary Note 3). After Fourier transforming the measured field distribution, we obtain nearly flat dispersions of the surface states around the projected K′ and K points, as shown in Fig. 2e, g, respectively. Those flat dispersions are characteristic signatures of drumhead surface states [21,25-32]. Measurements on the bottom boundary exhibit similar dispersions and thus are omitted here. The drumhead surface states emerge within the projected region of the nodal rings around the K and K′ points on the surface BZ, as manifested by calculated dispersions in Fig. 1j and Fig. 2e, g. In contrast to previously observed drumhead surface states [21,25-32], our design hosts only two such states at distinct frequencies (8.8 GHz and 7.0 GHz around the projected K and K′, respectively), a consequence of time-reversal and inversion symmetries breaking.

We then apply similar strategies to measure the front and back boundaries (parallel to the $xz$-plane). The measured surface dispersions at $k_z = 0$, $0.5\pi/h$, and $\pi/h$ slices for the front and back boundaries are presented in Fig. 2f, h, respectively. In both cases, the dispersions exhibit a strictly positive group velocity along the $+x$ direction, confirming the emergence of antichiral surface states. Theoretical calculations (Fig. 1l, white shaded region and cyan lines in Fig. 2f, h) show that antichiral surface states connect the two projected Dirac lines along K–H and K′–H′ with a large frequency shift, in agreement with the measurements. Distinct from the antichiral surface states in magnetic Weyl crystals [16] that host merely one transport channel on each surface, our system supports two surface sheets per boundary, which enables the construction of multimode one-way waveguides with additional co-propagating surface channels and gives rise to richer topological boundary states (including the antichiral hinge states shown below).

To explain the physical origin of the drumhead and antichiral surface states, we characterize the bulk topology using the Wilson loop formalism. The bulk polarization along a given direction is defined as [33]:

$$\nu_l^{\text{bulk}}(k_m,k_n)=\frac{1}{2\pi}\arg\det W_l(k_m,k_n)=\frac{1}{2\pi}\oint dk_l\,\mathrm{Tr}\,A_l(k_m,k_n)\quad(\mathrm{mod}\,1),\tag{1}$$

where $W_l(k_m,k_n)=\mathcal{P}\exp[i\oint dk_l A_l(k_m,k_n)]$ is the Wilson loop, $\mathcal{P}$ denotes path ordering, and $A_l(k_m,k_n)$ is the non-Abelian Berry connection evaluated for the lowest two bulk bands. The corresponding Wannier bulk polarization $\nu_z(k_x,k_y)$ is well defined at gapped momenta, where it takes a value of 1/2 inside the region enclosed by the nodal ring and 0 outside; it is undefined only on the nodal-ring contour itself, where the bulk gap closes, signaling the emergence of drumhead surface states when the system is finite along the $z$-axis. Similarly, $\nu_y(k_x,k_z)$, which is evaluated from the Wilson loop along the $k_y$ direction, characterizes the $y$-directed bulk polarization. In the momentum window between the projected nodal rings, this polarization remains nontrivial on the gapped Wilson-loop cycles, predicting antichiral surface states for open boundaries along the $y$ direction (see Supplementary Note 4). The singular cycles where the Wilson-loop path intersects the nodal rings are excluded from the polarization assignment (Supplementary Note 4). Furthermore, the measured transmission spectra of surface states confirm the unique transport signatures of these two surface modes: the drumhead surface states exhibit characteristic dual-frequency intensity peaks corresponding to the shifted nodal rings, whereas the antichiral surface states demonstrate robust nonreciprocal transmission consistent with our theoretical predictions (see Supplementary Note 3).

**Experimental observation of antichiral hinge states**

To observe the hinge states, we first fabricated a sample with an odd number of photonic layers (11 layers). All boundaries parallel to the $x$-$y$ and $x$-$z$ planes are covered with copper cladding, while the remaining boundaries (parallel to the $y$-$z$) are wrapped with microwave absorbers. As illustrated in Fig. 3a, a point source placed near hinge A excites fields propagating along the $+x$ direction, as captured by near-field scanning throughout the sample at 7.2 GHz. Frequency-dependent transmission spectra, obtained by positioning source antennas at both ends of hinge A, are plotted in Fig. 3b. Within the frequency range between the two nodal rings (7.0 to 8.8 GHz, shaded region in Fig. 3b), the forward transmission (purple solid line) exceeds the backward transmission (orange solid line) by up to approximately 30 dB. A Fourier transform of the measured near-fields along hinge A exhibits a dispersion with positive group velocity along $k_x$, indicative of chiral propagation (Fig. 3c). Repeating this procedure around hinge C also reveals unidirectional transport along the $+x$ direction, as summarized also in Fig. 3a–c (dotted purple and orange lines in Fig. 3b). Theoretical calculations show that a pair of degenerate hinge states connects the projected nodal rings at K and K′ (Fig. 1k and Fig. 3c), in good agreement with the measurements. These hinge arcs terminate near the projected nodal rings because the projected bulk gap closes there, causing the lateral antichiral surface bands to merge into the bulk continuum and rendering their surface polarization ill-defined. Although drumhead surface states also appear in the projected nodal-ring regions on the z-normal surfaces, they arise from a distinct bulk-boundary correspondence of the nodal rings and do not provide the topological origin of the hinge arcs. The robustness of these hinge states is further validated by introducing large metallic obstacles along their propagation paths; as shown in Fig. 3d–g, the waves bypass the obstacles without an obvious increase in reflected-field intensity in the measured maps.

Notably, both diagonal hinges (hinges A and C) support co-propagating chiral modes, manifesting the characteristic antichiral hinge states. This diagonal distribution is consistent with the surface-polarization criterion once the termination-dependent unit-cell convention is considered. For the odd-layer sample, the two-layer unit cells defined from the top and bottom boundaries differ by one layer; therefore, the surface polarization for the same lateral surface has 1/2 at one $z$-termination but 0 at the other. Consequently, the surface-polarization mismatch appears on opposite $y$-normal sides at the top and bottom boundaries, selecting the non-coplanar diagonal hinges A and C. Other hinges of the sample, including hinges B and D, exhibit no such unidirectional dispersive modes, demonstrating that the topological transport channels are spatially selective (see Supplementary Note 5). Note that a four-hinge configuration, with hinge states appearing simultaneously at hinges A, B, C, and D, can also be obtained by setting identical interlayer couplings for the two sublattices while alternating the couplings between adjacent layers (see Supplementary Note 5). A similar four-boundary response may also be mimicked by attaching two modified Haldane layers to the top and bottom boundaries of a 3D trivial insulator. In contrast, the selective emergence of the non-coplanar diagonal hinge states at hinges A and C originates from the surface-polarization mismatch of the intrinsically coupled 3D nodal-ring semimetal, and cannot be reduced to a passive assembly of independent 2D antichiral edge systems. Unlike chiral boundary states that can form closed cross-dimensional loops [17,18,34], the co-propagating nature of our antichiral hinge states prevents closed loops on a rectangular geometry; nevertheless, such loops can be realized by adopting a triangular-prism geometry (see Supplementary Note 5).

This origin of the antichiral hinge states can be intuitively understood by viewing the stacked 3D photonic crystal as a 1D Su-Schrieffer-Heeger (SSH)-like system [35]. The copper plates have two types of termination along the $y$-direction (see Fig. 2c, d), producing alternating interlayer couplings of the surface states on both the front and the back boundaries. The two alternating interlayer couplings function as the intracell and intercell hoppings in the SSH model. In this picture, the hinge states arise from the polarization of the antichiral surface states, analogous to the end states that emerge from the bulk polarization in the SSH model [35]. Specifically, a hinge state appears when the interlayer coupling between the outermost layer and its neighbor is less than that between the neighbor and the next layer. These phenomena are rigorously characterized by the quantized surface polarization $\nu_z^{\text{surface}}(k_x)$:

$$\nu_z^{\text{surface}}\left(k_x\right)=\frac{1}{2\pi}\oint dk_z A_z^{\text{surface}}\left(k_x,k_z\right)\ \left(\text{mod } 1\right) \tag{2}$$

where $A_z^{\text{surface}}(k_x,k_z)=i\langle u^{\text{surface}}(k_x,k_z)|\partial_{k_z}|u^{\text{surface}}(k_x,k_z)\rangle$ is the Abelian Berry connection evaluated over the antichiral surface band (see Supplementary Note 6). A nontrivial value of $\nu_z^{\text{surface}} = 1/2$ ($2/3 < k_x/(\pi/a) < 4/3$) predicts an antichiral hinge state at an interface where the termination-resolved surface polarization changes from 1/2 to 0. This criterion is valid only in the wavevector region where the antichiral surface band remains isolated from the projected bulk continuum. Outside this region, the surface branch merges into the projected bulk continuum, so the isolated surface band—and hence its surface polarization—cannot be defined. Within this well-defined momentum domain, this termination-dependent surface-polarization criterion identifies our system as a higher-order topological hierarchy, in which the one-dimensional hinge response is governed by the topology of the two-dimensional antichiral surface band. Related hierarchical boundary correspondence has been established in Dirac hierarchy [36,37], boundary-/surface-obstructed topology [38,39], edge-corner correspondence [40], and surface-polarization-induced hinge states [41].

**Spatial reconfiguration of the antichiral hinge states**

To reconfigure the hinge states, we attach one additional photonic layer to the top boundary of the sample, as illustrated in Fig. 4a. This added layer itself is a Dirac semimetal that hosts a pair of antichiral edge states [13,24]. Interestingly, upon attachment, the coupling between the antichiral edge state of a Dirac semimetal and the antichiral hinge state of a nodal ring semimetal at hinge C results in the disappearance of the hinge-localized state at hinge C, while a new hinge state emerges at hinge B. The reconfiguration is not a cancellation of co-propagating modes, but a change of the topological boundary termination: the added layer removes the surface-polarization mismatch at hinge C and creates a new one at hinge B. Both hinges A and B now support hinge states propagating with the same positive group velocity along the $+x$ direction, as verified by the theoretical calculations and experimental measurements shown in Fig. 4b–d. This tunability of hinge-state position originates from the change in surface polarization (see Supplementary Note 6). Although the present implementation is modular, it does not require refabrication and could be further extended to dynamic control by mechanically tuning the layer spacing or by using controllable shutters at the interlayer coupling holes. Furthermore, the antichiral hinge channels can be deterministically distributed across multiple step edges in stepped geometries, strictly depending on the even/odd layer parity of the truncation (see Supplementary Note 5).

**Discussion**

In summary, we have theoretically proposed and experimentally realized a 3D higher-order photonic nodal ring semimetal. We unveil a topological cascade that connects gapless nodal rings and Dirac lines in the bulk to flat drumhead surface states, dispersive antichiral surface states, and ultimately to antichiral hinge states. Furthermore, we show that the spatial position of these hinge channels can be deterministically reconfigured purely by adding or removing photonic layers, enabling the on-demand control of antichiral hinge channels. All the topological boundary states are characterized and unified by our calculated bulk and surface polarization. Our platform not only extends antichiral transport to 3D higher-order topologies, but also provides a versatile platform for exploring the interplay between gapless bulk singularities and multidimensional topological boundary responses. By combining antichiral surface transport with switchable antichiral hinge channels, it further offers a route to high-capacity 3D topological coaxial-cable-like photonic devices for programmable non-reciprocal routing.

## Methods

### Theoretical model and physical mechanism

The 3D photonic nodal ring semimetal is theoretically described by a four-band tight-binding model defined on a periodic stack of honeycomb bilayers. In the Bloch basis $\left(c_{A1,k}, c_{B1,k}, c_{A2,k}, c_{B2,k}\right)^T$, where A/B denote the sublattices and 1/2 label the two layers within a unit cell, the Bloch Hamiltonian $H(k)$ is written as:

$$H(\mathbf{k}) = \begin{pmatrix} H_1(\mathbf{k}_{||}) & V(k_z) \\ V^{\dagger}(k_z) & H_2(\mathbf{k}_{||}) \end{pmatrix} \tag{3}$$

Here, the diagonal blocks $H_1(\mathbf{k}_{||})$ and $H_2(\mathbf{k}_{||})$ represent the single-layer modified Haldane models, which break time-reversal symmetry to support antichiral properties without opening a valley bandgap. The off-diagonal block describes the critical interlayer coupling:

$$V(k_z) = \begin{pmatrix} t_A(k_z) & 0 \\ 0 & t_B(k_z) \end{pmatrix} \tag{4}$$

where $t_A(k_z)$ and $t_B(k_z)$ are the vertical hopping amplitudes for sublattices $A$ and $B$, respectively. Instead of a trivial extension, our system introduces a staggered interlayer coupling mechanism where the weak and strong couplings alternate. This specific spatial arrangement shifts the energy of the Dirac cones, leading to the formation of continuous nodal rings in the bulk momentum space. Crucially, this staggered configuration generates a rigorous surface-polarization mismatch between different boundaries, which macroscopically determines the dimensional hierarchy and ultimately dictates the emergence of the 1D antichiral hinge states. Full mathematical derivations of the Hamiltonian are provided in the Supplementary Information.

### Numerical simulations

Full-wave numerical simulations were performed using the RF module of COMSOL Multiphysics. The perforated copper plates and permanent magnets were modeled with Perfect Electric Conductor (PEC) boundaries. The Yttrium Iron Garnet (YIG) rods were modeled as a gyromagnetic material characterized by a frequency-dependent Polder permeability tensor. Under an external static magnetic bias applied along the $z$-axis, the tensor takes the form:

$$\hat{\mu} = \begin{pmatrix} \mu_r & i\kappa & 0 \\ -i\kappa & \mu_r & 0 \\ 0 & 0 & 1 \end{pmatrix} \tag{5}$$

where the diagonal and off-diagonal components are given by $\mu_r = 1 + \frac{\omega_m \omega_0}{\omega_0^2 - \omega^2}$ and $\kappa = \frac{\omega_m \omega}{\omega_0^2 - \omega^2}$, respectively. Here, $\omega_0 = \gamma H_0$ is the Larmor precession frequency, $\omega_m = \mu_0 \gamma M_s$, and $\gamma = 1.76 \times 10^{11}$ C/kg is the gyromagnetic ratio. In our simulations, we utilized a saturation magnetization of $4\pi M_s = 1780$ Gauss and a bias field of $H_0 = 180$ mT, strictly consistent with our experimental configurations. The dispersion relations for the topological hierarchy were calculated using the eigenfrequency solver with dimension-dependent boundary conditions: (i) Bulk states: Periodic boundary conditions (PBC) were applied along all three directions ($x$, $y$, $z$). (ii) Drumhead surface states: PBC were applied along the $x$- and $y$-directions, with PEC boundaries along the $z$-direction. (iii) Antichiral surface states: PBC were applied along the $x$- and $z$-directions, with PEC boundaries along the $y$-direction. (iv) Antichiral hinge states: PBC were applied only along the propagation direction ($x$), while PEC boundaries were imposed along both transverse directions ($y$ and $z$). For the transmission spectra and spatial field distributions, we employed the frequency domain solver, utilizing Scattering Boundary Conditions (SBC) at the source and drain ends to minimize undesired reflections.

**Sample fabrication and experimental measurements**

The metallic layers of our 3D photonic crystal were constructed from mechanically machined copper plates. The magnetic components consist of YIG rods stacked with permanent magnets, arranged in a honeycomb lattice. The permanent magnets provide a static bias magnetic field of approximately 180 mT, which is sufficient to fully saturate the YIG rods and break time-reversal symmetry. To support and precisely fix these components, we used perforated foam boards (ROHACELL 31 HF) with a relative permittivity of 1.04 and a low loss tangent of 0.0025, matching the thickness of the magnetic-YIG combination layers. To map the topological modes, the microwave near-field measurements were performed using a vector network analyzer (Keysight E5063A) connected to a computer-controlled three-dimensional scanning stage. The source and probe antennas were semi-rigid coaxial cables with an outer diameter of 1.2 mm. During the measurements, the antennas were inserted into the central circular air holes of the unit cells to excite and detect the local electric field. These air holes have a diameter of 1.6 mm, allowing reproducible insertion of the coaxial cables without damaging the sample or changing the lattice geometry. Since the field was sampled at the center of each unit cell, the in-plane scanning step and spatial sampling resolution were set by the lattice constant $a$ = 10 mm; along the vertical direction, the field maps were recorded layer by layer. For the hinge-transport measurements, the source was placed at the outermost air hole adjacent to the selected hinge, and the sample contained 20 periods along the $x$ direction, corresponding to a total transmission length of 200 mm. To examine robustness against defects, metallic obstacles were introduced by inserting four metal pins into adjacent air holes along the propagation path, locally blocking the field and producing the unmeasured rectangular regions in the field maps shown in Fig. 3.

**Data availability**

The data that support the findings of this study are provided in the Supplementary Information and Source Data file. Source data are provided with this paper.

**Code availability**

The codes used in this paper are available from the corresponding authors upon request.

**Acknowledgements**

G.-G. Liu acknowledges funding from the National Natural Science Foundation of China (Grant No. 62575245), the Research Center for Industries of the Future (RCIF) at Westlake University (Grant No. 210000006022312), and the Westlake Education Foundation (Grant No. 103110736022301). Y. Yang acknowledges funding from the National Natural Science Foundation of China (Grant No. 12574477). S. Mandal and D. Mondal acknowledge funding support from the Industrial Research and Consultancy Centre (IRCC) at the Indian Institute of Technology Bombay (Seed Grant No. RD/0525-IRCCSH0-006). Y. Peng acknowledges funding from the China Postdoctoral Science Foundation (Grant No. 2025M773386). W. Qi acknowledges funding from the Young Scientists Fund of the National Natural Science Foundation of China (Grant No. 12104135) and the China Postdoctoral Science Foundation (Grant No. 2022M721057). R. Banerjee acknowledges support from the Anusandhan National Research Foundation (ANRF) (Grant No. RJF/2025/000278). B. Zhang acknowledges funding from Singapore National Research Foundation Competitive Research Program (Grant No. NRF-CRP23-2019–0007) and the Singapore Ministry of Education Academic Research Fund Tier 2 (Grant No. MOE-T2EP50123-0007).

**Author contributions**

G.-G.L. conceived the idea. Y.P. performed the simulation, designed the experiments, and analyzed the data. D.M., S.M., R.B., G.-G.L., and Y.P. performed the theoretical explanation and developed the tight-binding model simulations. Y.P., B.Z., S.W., Z.L., W.Q., X.D., M.C., W.L., X.Z., and F.-F.L. fabricated samples. W.W. and Y.Y. carried out the measurements. M.W., J.T., and P.Z. also discussed the results and contributed to the writing of the manuscript. Y.Y., S.M., B.Z. and G.-G.L. supervised the project.

**Competing interests**

The authors declare no competing interests.

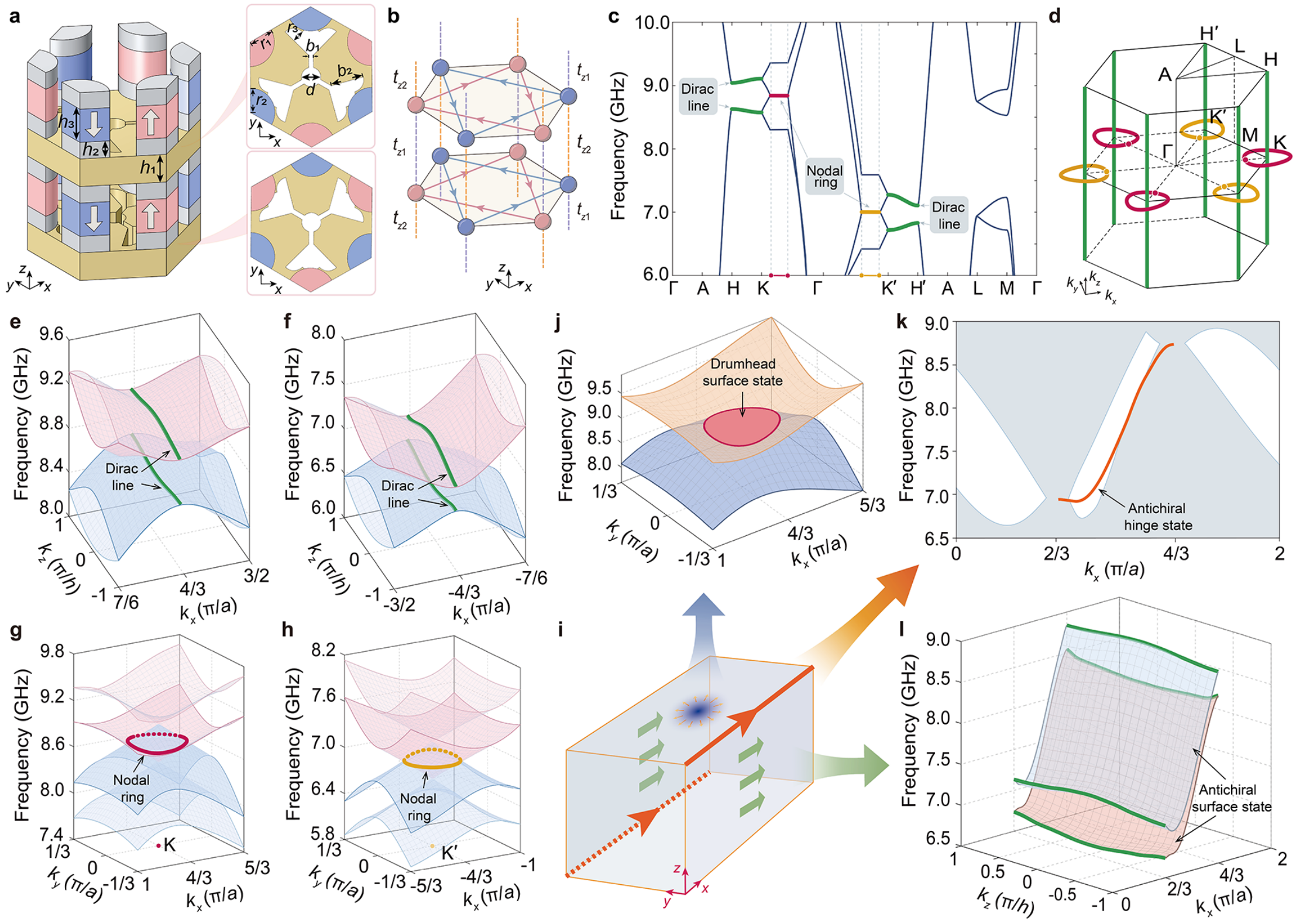

**Fig. 1 | Design of higher-order photonic nodal ring semimetal. a** Unit cell of the photonic crystal. Here $a$ = 10 mm, $h_1$ = 2.5 mm, $r_3$ = 1 mm, $d$ = 1.6 mm, $r_1$ = 2 mm, $h_3$ = 3 mm and $h_2$ = 1.5 mm. Right panels show the top views of the two copper plates. **b** Unit cell of the tight-binding model. **c** Calculated bulk band structure. The red and yellow lines on the *x*-axis indicate a path along the trajectory of the nodal rings. **d** 3D Brillouin zone of the photonic crystal. Green lines represent Dirac lines. Pink and yellow circles indicate nodal rings with shifted frequencies. **e**, **f** Bulk dispersion around the Dirac lines along the K–H and K′–H′, respectively. **g**, **h** Bulk dispersion around the nodal rings near the K and K′, respectively. **i–l** Illustration of a hierarchy of topological boundary states including drumhead surface states, antichiral surface states, and antichiral hinge states.

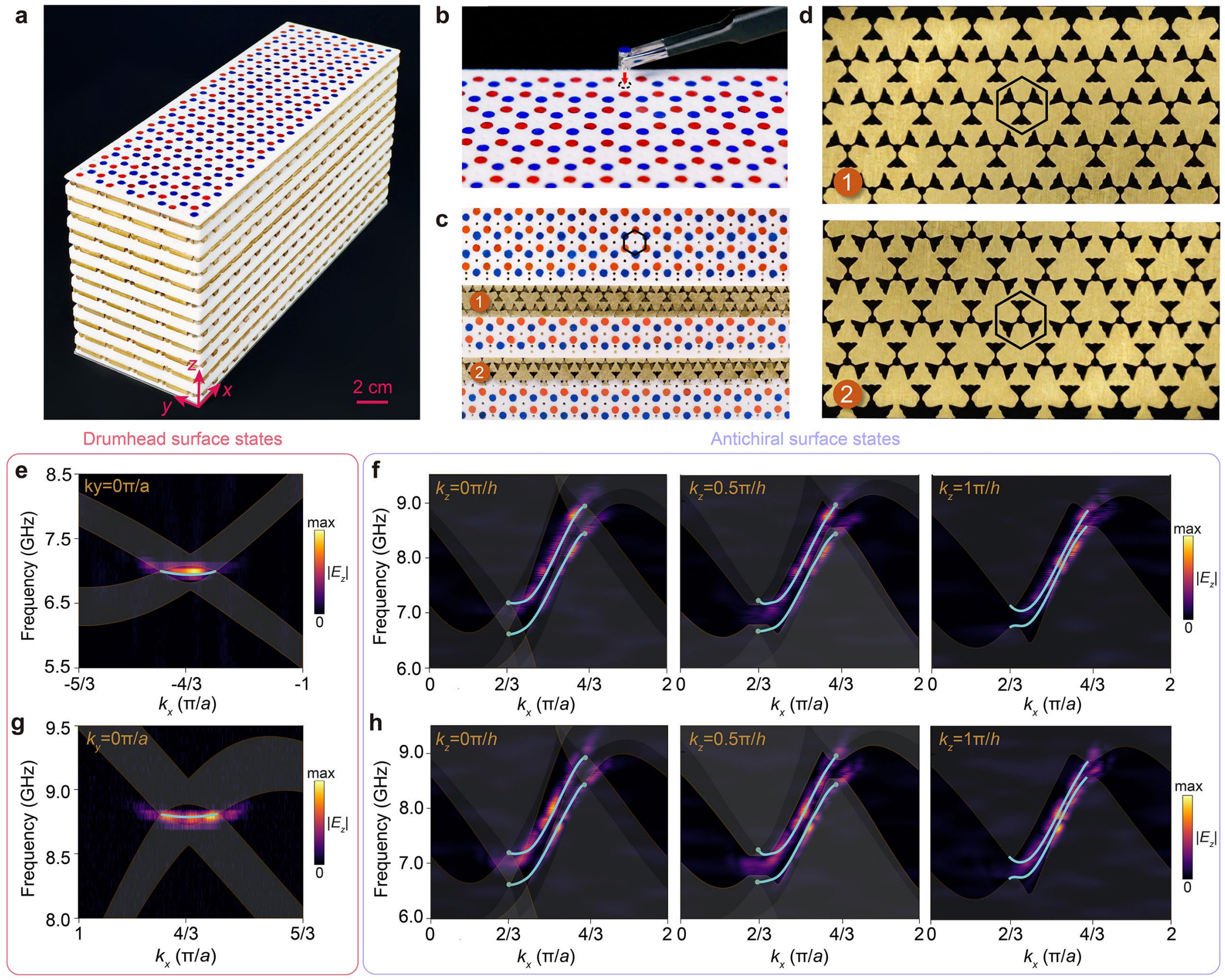

**Fig. 2 | Experimental observation of the surface states. a** Photograph of the fabricated photonic crystal sample. **b** Assembly process showing the insertion of YIG rods and permanent magnets. **c** Top view of the stacked structure with a shift between layers. **d** Zoomed-in photographs of the copper plates with distinct boundary geometries. **e**, **g** Measured dispersions of the drumhead surface state at the $k_y = 0$ plane near the projected K′ (**e**) and K (**g**). **f**, **h** Measured dispersions of the antichiral surface state for the front (**f**) and back (**h**) boundaries at $k_z = 0$, $0.5\pi/h$, and $\pi/h$. The cyan lines and the white shaded region represent the calculated surface states and projected bulk state, respectively.

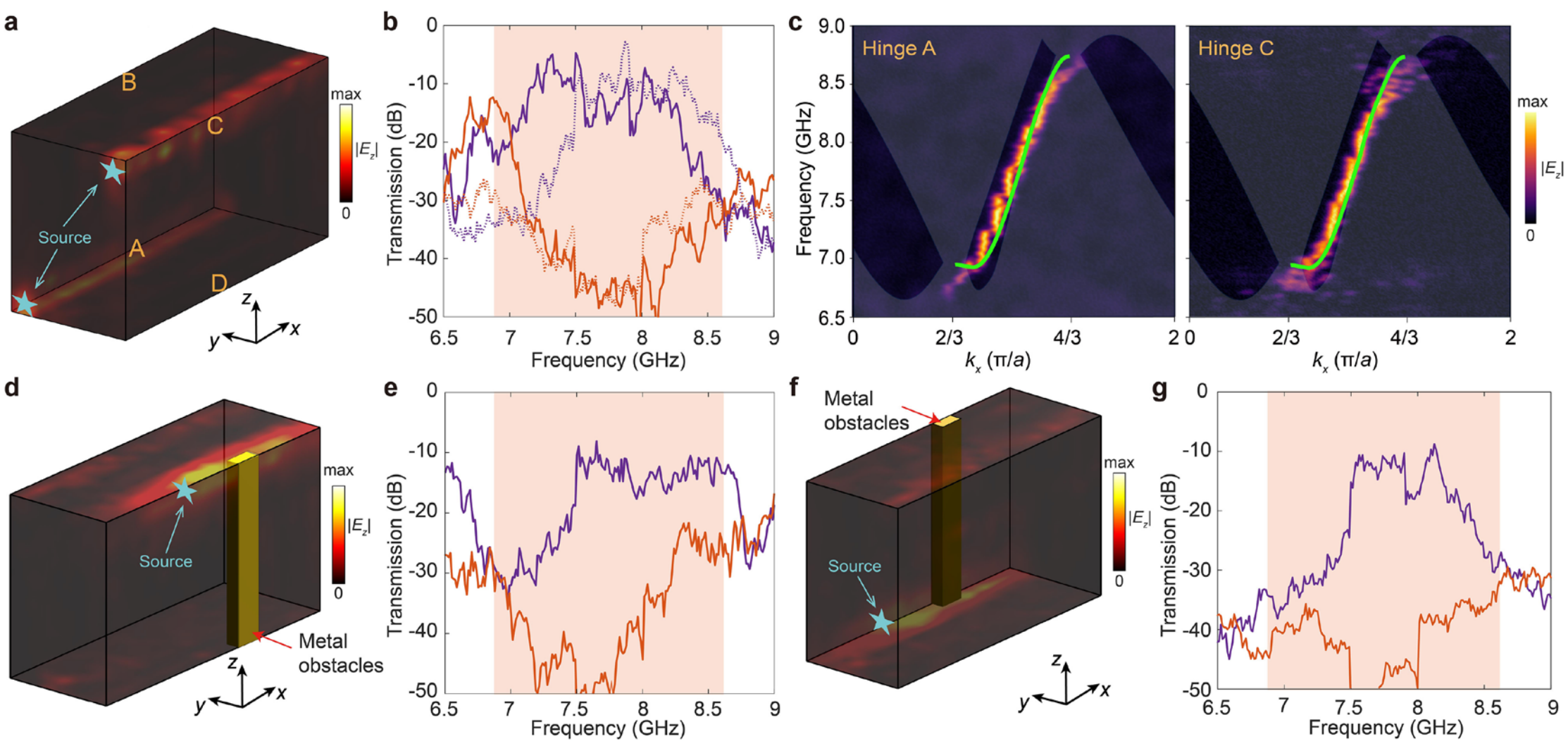

**Fig. 3 | Observation of antichiral hinge states. a** Measured field distribution around the hinges A and C. Blue stars indicate the source positions. **b** Measured transmission spectra for hinges A (solid lines) and C (dotted lines). Purple and orange lines are measured forward and backward transmission spectra, respectively. Orange shaded region is the frequency range between two nodal rings (7.0–8.8 GHz). **c** Measured dispersions of the hinge states around hinges A and C. The green lines and the white shaded region represent the calculated hinge states and projected bulk state, respectively. **d** Measured field distribution around the hinge C with a metallic obstacle placed near it. **e** Measured transmission spectra for hinge C. **f**, **g** Similar to **d**, **e** but for hinge A.

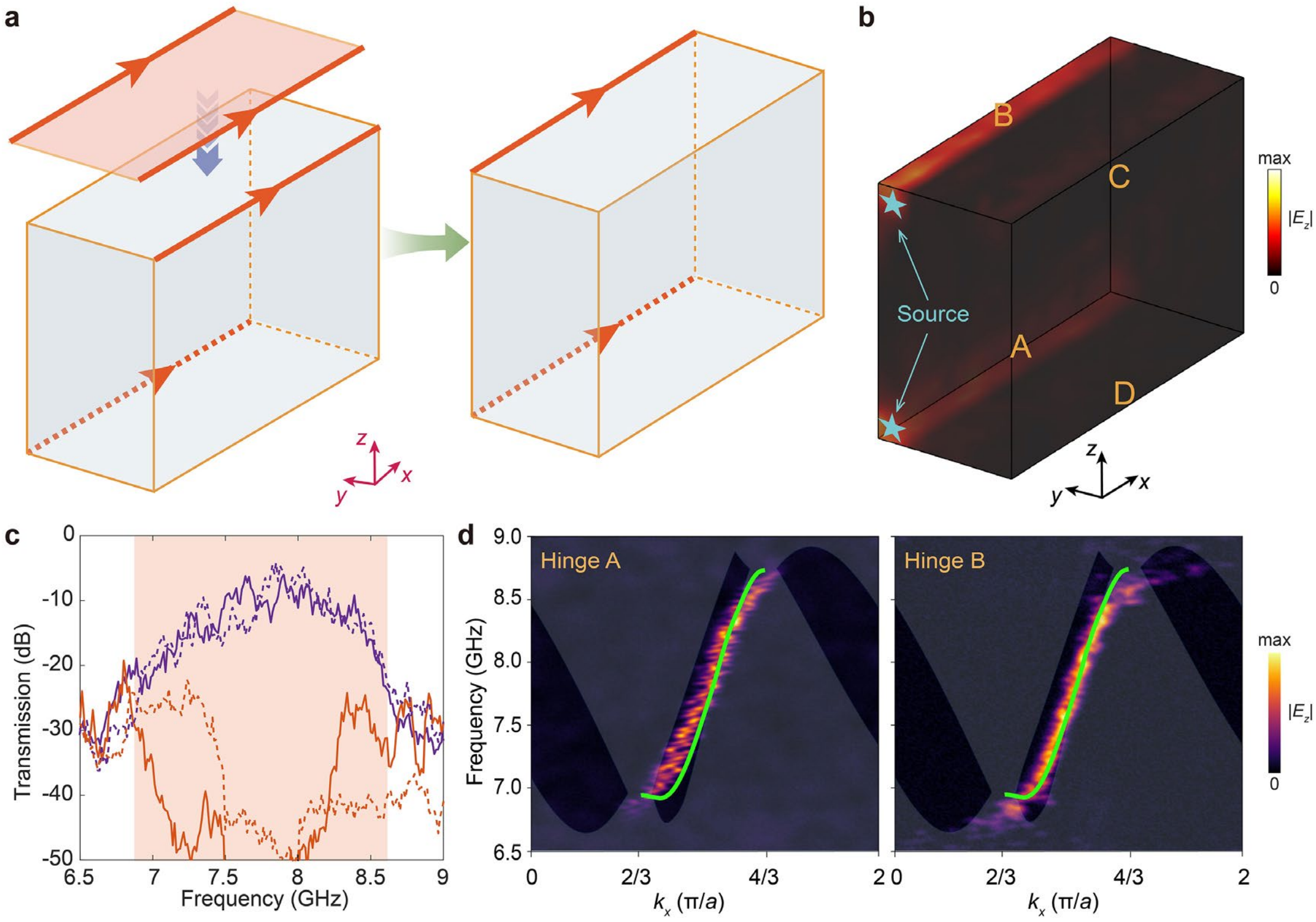

**Fig. 4 | Reconfiguration of antichiral hinge states. a** Schematic illustration of the spatial manipulation of hinge states. **b** Measured field distribution around the hinges A and B. Blue stars indicate the source positions. **c** Measured transmission spectra for hinges A (solid lines) and B (dotted lines). Purple and orange lines are measured forward and backward transmission spectra, respectively. Orange shaded region is the frequency range between two nodal rings (7.0–8.8 GHz). **d** Measured dispersions of the hinge states around hinges A and B. The green lines and the white shaded region represent the calculated hinge states and projected bulk state, respectively.